\documentclass[ngerman,english]{bvm} 

\bvmdef\articlenumber{3056}

\bvmdef\type{V}

\date{}

\title{Learning-Based Patch-Wise Metal Segmentation with Consistency Check}

\titlerunning{Patch-Wise Metal Segmentation with Consistency Check}

\author{Tristan~M.~Gottschalk$^{1,2,3}$, Andreas~Maier$^{1,3,4}$, Florian~Kordon$^{1,2,3}$, Bj\"orn~W.~Kreher$^{2}$}

\authorrunning{Gottschalk et al.}

\institute{%
$^1$Pattern Recognition Lab, Universit\"at Erlangen-N\"urnberg (FAU), Erlangen\\
$^2$Siemens Healthcare GmbH, Forchheim\\
$^3$Erlangen Graduate School in Advanced Optical Technologies (SAOT), Universit\"at Erlangen-N\"urnberg (FAU), Erlangen\\
$^4$Machine Intelligence, Universit\"at Erlangen-N\"urnberg (FAU), Erlangen}

\email{Tristan.Gottschalk@fau.de}

\begin{document}

%
\selectlanguage{english}

\maketitle

\begin{abstract}
Metal implants that are inserted into the patient's body during trauma interventions cause heavy artifacts in 3D X-ray acquisitions. Metal Artifact Reduction (MAR) methods, whose first step is always a segmentation of the present metal objects, try to remove these artifacts. Thereby, the segmentation is a crucial task which has strong influence on the MAR's outcome. This study proposes and evaluates a learning-based patch-wise segmentation network and a newly proposed Consistency Check as post-processing step. The combination of the learned segmentation and Consistency Check reaches a high segmentation performance with an average IoU score of 0.924 on the test set. Furthermore, the Consistency Check proves the ability to significantly reduce false positive segmentations whilst simultaneously ensuring consistent segmentations.
\end{abstract}

\section{Introduction}
Trauma interventions regularly require an intraoperative evaluation of the correct positioning of inserted metallic implants. This oftentimes not only includes a 2D- but also a 3D-X-ray scan performed by a mobile C-arm. Due to heavy image artifacts caused by the inserted implants the surgeon is in need of a well functioning metal artifact reduction method (MAR). Thereby, the overall ability of the particular MAR method to reduce the artifacts heavily depends on the quality of the so-called metal segmentation \cite{3056-01}. Experiments show that missed metal parts in the segmentation lead to still present streak artifacts in the reconstructions even after the MAR, whereas falsely segmented anatomical structures lead to a blurred or even vanished representation. Such segmentation can be done in the 2D projections as well as in the corresponding 3D reconstructed volume. Whereas classic MAR methods like normalized and frequency split metal artifact reduction method (NMAR,FSMAR) \cite{3056-02,3056-03} use advanced threshold-based metal segmentation methods in 3D, recent segmentation methods has shown the high potential of deep learning based models like e.g. shown by Ronneberger et al. \cite{3056-04}. Despite these advancements, recent MAR methods oftentimes still use thresholding methods in 3D for metal segmentation in the volume \cite{3056-05,3056-06} although this method has clear disadvantages. Firstly, the present artifacts in the 3D volume (caused by the metal itself) strongly aggravates the segmentation task and makes a clean thresholding difficult and secondly, metal which lies outside the field of view (FOV) of the reconstructed volume can not be segmented at all. Thus, the unsegmented metal parts will not be processed by the MAR and consequently still cause heavy artifacts in the volume. In contrast, segmentation methods on the 2D projection images have the possibility to find all present metal and thus should be preferred.

In order to tackle the mentioned problems, we investigate the difference in performance of 2D projection-based networks, the first using a patch-wise segmentation which is trained and tested for two patch-sizes and using a sliding window with stitching and the second performing segmentation on the complete projection image at once. Furthermore, a segmentation Consistency Check (CC) is introduced as a post processing step, which robustly removes falsely segmented structures, whilst ensuring consistent segmentation masks.

\section{Materials and methods}
\subsection{Network architectures}
Due to the proven performance of the U-Net architecture in segmenting medical images \cite{3056-04}, the proposed networks are inspired by it. Thereby, the non-patched architecture version which has an input image size of 976x976 pixels, consists of seven contracting blocks and seven expanding blocks with skip-connections in between. Starting with 8 feature maps in its first layer, the network has 1024 feature maps at its bottleneck, each with a size of 8x8 pixels. The expanding path of the network is designed that it expands the processed image to its initial size. In accordance with that, its patch-based counterpart has input image sizes of 128x128 and 256x256 pixels and the same-/doubled-sized bottleneck. Thus, it consists of only 4 contracting and 4 expanding blocks.

\subsection{Data}
To train the networks, we acquired suitable X-ray projection images by performing two consecutive 3D short-scans, the first with and the second without metal implants.
These scans were collected during two cadaver studies of human knees, using a Siemens Cios Spin system and following an acquisition protocol similar to \cite{3056-07}. During both studies, in total 32 corresponding 3D scans were acquired, each consisting of 400 projection images, covering a variety of metal implants, that were placed on the skin or directly on and inside the bone. The corresponding ground truth (GT) labels for the metal segmentation task can be generated by subtraction. Thus, we are able to train our networks using real acquired X-ray projections which should provide a good generalization to clinical knee data. The data set was split into 22 3D scans for training, 8 for testing and 2 for validation.

\subsection{Experiment protocol}
As a preprocessing step, the acquired RAW data was converted into line integral data. This was done by using Lambert-Beer Law \cite{3056-08}, where the measured intensities were normalized with the initially emitted intensity and then the logarithm was applied.
Both networks were trained from scratch using the same data split, the same cross-entropy based loss function, a decaying learning rate (start: $1e^{-6}$) and an online augmentation scheme with randomized left-right flips, rotations, contrast and brightness scaling and different amount of added Poisson noise to mimic varying image acquisition qualities. For training of the patch-based segmentation networks, a randomized cropping with sizes of 128 or 256, resp. is added to the online augmentation scheme. All networks were trained until convergence. In cases of patch-wise segmentation, a sliding window (step size 32 or 64 resp.) with stitching was implemented, where the values of the overlapping windows are summed up in order to segment the complete projection.

\subsection{Consistency check}
By exploiting the underlying consistency conditions of 3D reconstructions, we propose a segmentation Consistency Check which is able to reduce false positive segmentations and simultaneously accounts for consistency of the mask among the stack of segmentations and which consists of three steps. The first step is a back-projection of the complete stack of binarized segmentation masks into a 3D volume. In contrast to the initial diagnostic volume which has in our case 512$^3$ voxels (voxel size of \(0.31\,\textrm{mm}\)), the size of this volume is chosen larger and in a way that the reconstruction of the binarized masks includes all metal parts present. This leads to a volume size of 920$^3$ voxels with again a voxel size of \(0.31\,\textrm{mm}\). Due to the fact that we back-project binary masks, each voxel of the corresponding volume can be understood as a ``visitor counter'' which provides information about how many of the 2D metal masks actually contributed to each specific voxel. By normalizing each voxel value with the maximal amount of possible ``visitors'', we account for the reduced amount of rays outside the FOV of the volume. Consequently, the voxel values lie in the range of 0 and 1, where a voxel with a value of 0 corresponds to no contribution and a voxel with a value of 1 corresponds to a contribution from all 2D metal masks. Thus, each voxel value provides information about how consistently the corresponding metal part was segmented across the projections. Consequently, applying a threshold of e.g. 0.95 in order to create a binary 3D metal mask in the reconstruction, is equivalent to only including the parts into the 3D mask that were segmented in at least 95\% of the projections. Thus, by performing the final forward projection, a stack of 2D metal masks that only include consistently segmented parts, is provided. Consequently, the inconsistent false positive segmentations are removed. Thereby, the classification into consistent or inconsistent parts is defined by the chosen threshold.

\section{Results}
\begin{table}[tb]
\caption{Results for the AUC of the three different segmentation networks.The results are calculated over the complete test data set.}
\label{3056-tab-01}
\centering
\begin{tabular*}{0.5\textwidth}{l@{\extracolsep\fill}llll}
\hline
Patch Size              & Avg.       & Min       & Max \\ \hline
No patching   & 0.967     & 0.952     & 0.990 \\
256x256                 & 0.972     & 0.960     & 0.984 \\
128x128                 & \textbf{0.983}     & \textbf{0.973}     & \textbf{0.994} \\
 \hline
\end{tabular*}
\end{table}

By having a look at Tab. \ref{3056-tab-01}, it can be observed that all three networks perform on a high level. Whereas the patch-based segmentation using the patch-size of 128 results in the largest AUC, the increased patch-size of 256 results in a slightly decreased AUC. Using the network segmenting on the complete projection, generates the lowest AUC.

\begin{table}[b]
\centering
\begin{tabular*}{\textwidth}{l@{\extracolsep\fill}llllll}
\hline
Thres. & CC  & Avg. IoU & Avg. Dice & Avg. Precision & Avg. Recall \\ \hline
5  & no     & 0.783$\pm$0.107     & 0.855$\pm$0.095 & 0.598$\pm$0.200  & 0.995$\pm$0.023  \\
5  & yes    & 0.924$\pm$0.032     & 0.959$\pm$0.024 & 0.873$\pm$0.039  & 0.975$\pm$0.038  \\
30 & no     & 0.913$\pm$0.034     & 0.952$\pm$0.026 & 0.850$\pm$0.052  & 0.978$\pm$0.041  \\
30 & yes    & 0.917$\pm$0.042     & 0.954$\pm$0.029 & 0.917$\pm$0.031  & 0.908$\pm$0.073  \\
55 & no     & 0.913$\pm$0.040     & 0.952$\pm$0.029 & 0.895$\pm$0.042  & 0.925$\pm$0.076  \\
55 & yes    & 0.832$\pm$0.084     & 0.954$\pm$0.060 & 0.954$\pm$0.027  & 0.707$\pm$0.171 \\
 \hline
\end{tabular*}
\caption{Average results of the 128 patch-sized network for IoU, Dice Score, Precision and Recall for different thresholds for mask binarization in combination with or without the proposed CC.}
\label{3056-tab-02}
\end{table}

\begin{figure}[b]
	\setlength{\figbreite}{0.99\textwidth}
	\centering
	\includegraphics[width=\figbreite]{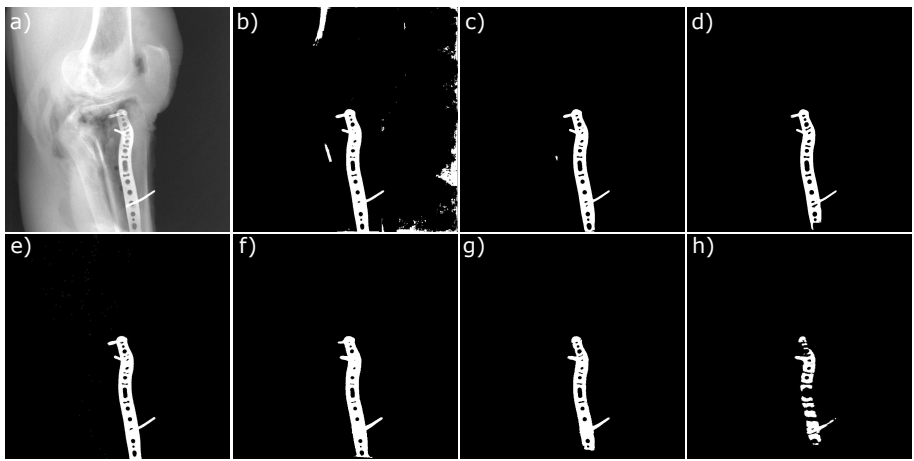}
	\caption{a) and e) show the 2D projection and GT mask, whereas b),c) and d) correspond to the binary result masks for the thresholds 5, 30 and 55 using the 128 patch-sized network. Images f) to h) show the masks after being processed by the proposed CC.}
	\label{3056-fig-01}
\end{figure}

In Tab. \ref{3056-tab-02} the results for the combination of three different thresholds (5, 30, 55) for the binarization of the patch-wise network's output with and without the proposed CC are illustrated. It can be seen that using the thresholds 30 and 55 for binarization without CC result in overall high metric scores, whereas using a threshold of 5 results in significant lower scores for all metrics, expect the avg. recall. The shown images b) to d) of Fig. \ref{3056-fig-01} support the quantitative results. It can be seen that the mask in image b) contains a higher amount of falsely segmented structures. However, especially at the bottom edge, the metal implant is still segmented, whereas those parts are missing in images c) and d).
Having a look at the configurations applying the proposed CC, it can be observed that the evaluation metrics are significantly increased when being combined with the lowest threshold of 5. This combination reaches the highest avg. IoU and Dice score among all tested configurations (cf. Tab. \ref{3056-tab-02}). Whereas the scores marginally increase for the threshold 30 with CC, decreasing scores can be investigated for threshold 55 with CC. The observed effects are also visible in images f) to h) in Fig. \ref{3056-fig-01}.

\section{Discussion}
By investigating the AUC values for the three evaluated network setups, it becomes clear that all setups perform on a high level with mean AUC values of almost reaching 1. Having a closer look shows that patching in general seems to be beneficial and that a reduction of the patch-size further increases the performance. However, in future work, the influence of more patch-sizes should be investigated. Based on these results the subsequent tests of the proposed CC were evaluated using the best-performing network with a patch size of 128.

When neglecting the CC at first, it can be investigated that using thresholds 30 and 55 result in an supposedly high performance with respect to the quantitative results (cf. Tab. \ref{3056-tab-02}), whereas using the threshold 5 leads to a poor performance. This effect is due to the high amount of falsely segmented structures as can be seen in Fig. \ref{3056-fig-01} image b). However, despite the high metric scores and rather convenient looking mask for threshold 30 and 55, neither the quantitative nor the qualitative evaluation provide distinct information about how consistent the respective segmentations are. Due to the fact that consistency among the complete stack of segmented masks of a 3D scan is crucial for the quality of the corresponding reconstruction, this information is important. Inconsistencies in the segmented metal masks will lead to streaking artifacts in the volume. Having a look at the results after applying the proposed CC, it can be observed that the performance of the threshold 5 is significantly boosted with respect to the quantitative results, whereas the performance of the threshold 30 only marginally increases and the performance for the threshold of 55 even decreases. The distinct increase for the threshold of 5 can be explained by the complete removal of false positive segmentations. The same holds for threshold 30. However, a more interesting result can be drawn from image h) in Fig. \ref{3056-fig-01}. Applying the CC reveals and simultaneously fixes the problem that the initial segmentation using a threshold of 55 was inconsistent which results in removed metal parts. Despite the correspondingly decreased segmentation metrics, we are confident that the newly reached consistency within the mask will result in less artifacts in the corresponding volume. However this hypothesis needs to be further investigated. Concluding the discussion about the proposed Consistency Check, it can be stated that CC is able to remove falsely positive segmented structures whilst simultaneously enforcing consistency of the segmentation. Both should be beneficial for subsequent MAR methods. Based on the fact that the proposed patch-wise segmentation networks were trained using real X-ray scans, we are confident that the method generalizes well to clinical cases. Nonetheless, this should be further investigated in future work.

\ack
The authors gratefully acknowledge funding of the Erlangen Graduate School in Advanced Optical Technologies (SAOT) by the Bavarian State Ministry for Science and Art. Furthermore, the authors like to thank the Rimasys GmbH for their extensive support during the cadaver studies.

\dis
The methods and information presented here are based on research and are not commercially available.

\bibliographystyle{bvm}

\bibliography{3056}

\marginpar{\color{white}E\articlenumber}

\end{document}